\documentclass[10pt,conference]{IEEEtran} 
\IEEEoverridecommandlockouts
\usepackage{cite}
\usepackage{amsmath,amssymb,amsfonts}
\usepackage{algorithmic}
\usepackage{graphicx}
\usepackage{amsmath,amsfonts}
\usepackage{algorithm}
\usepackage{array}
\usepackage[caption=false,font=normalsize,labelfont=sf,textfont=sf]{subfig}
\usepackage{textcomp}
\usepackage{stfloats}
\usepackage{url}
\usepackage{verbatim}
\usepackage{graphicx}
\usepackage{cite}
\usepackage{xcolor}
\usepackage{bm}
\usepackage{multirow}
\usepackage{multicol}
\usepackage{textcomp}
\usepackage{xcolor}
\usepackage{booktabs}
\usepackage{hyperref}

\newif\ifcolormarker
\colormarkertrue

\def\BibTeX{{\rm B\kern-.05em{\sc i\kern-.025em b}\kern-.08em
    T\kern-.1667em\lower.7ex\hbox{E}\kern-.125emX}}
\newcommand{\NAME}{RBB}
\begin{document}

\title{Fast Traffic Engineering by Gradient Descent with Learned Differentiable Routing}

\author{\IEEEauthorblockN{Krzysztof~Rusek\IEEEauthorrefmark{2},
 Paul~Almasan\IEEEauthorrefmark{1},
José~Suárez-Varela\IEEEauthorrefmark{1},
Piotr Cho\l{}da\IEEEauthorrefmark{2},
Pere~Barlet-Ros\IEEEauthorrefmark{1}, \\
Albert~Cabellos-Aparicio\IEEEauthorrefmark{1}}
\IEEEauthorblockA{\IEEEauthorrefmark{2}AGH University of Science and Technology, Institute of Telecommunications\\
}
\IEEEauthorblockA{\IEEEauthorrefmark{1}Barcelona Neural Networking Center, 
Universitat Politècnica de Catalunya, Spain}
\thanks{\textbf{NOTE:} This work has been accepted for presentation in the International Conference on Network and Service Management (CNSM). © 2022 IEEE. Personal use of this material is permitted. Permission from IEEE must be obtained for all other uses, in any current or future media, including reprinting/republishing this material for advertising or promotional purposes, creating new collective works, for resale or redistribution to servers or lists, or reuse of any copyrighted component of this work in other works.}}

\maketitle

\begin{abstract}

Emerging applications such as the metaverse, telesurgery or cloud computing require increasingly complex operational demands on networks (e.g., ultra-reliable low latency). Likewise, the ever-faster traffic dynamics will demand network control mechanisms that can operate at short timescales (e.g., sub-minute). In this context, Traffic Engineering (TE) is a key component to efficiently control network traffic according to some performance goals (e.g., minimize network congestion).

This paper presents \textit{Routing By Backprop} (RBB), a novel TE method based on Graph Neural Networks (GNN) and differentiable programming. Thanks to its internal GNN model, RBB builds an end-to-end differentiable function of the target TE problem (\textit{MinMaxLoad}). This enables fast TE optimization via gradient descent. In our evaluation, we show the potential of RBB to optimize OSPF-based routing ($\approx$25\% of improvement with respect to default OSPF configurations). Moreover, we test the potential of RBB as an initializer of computationally-intensive TE solvers. The experimental results show promising prospects for accelerating this type of solvers and achieving efficient online TE optimization.
\end{abstract}


\section{Introduction}
Communication networks are witnessing a fast-paced transformation driven by the continuous emergence of new services and applications. Novel trends such as the metaverse (AR/VR), cloud computing, vehicular networks, or telesurgery require increasingly complex operational demands. This puts great pressure on network operators, which need to count on more advanced network mechanisms that can quickly react to the fast network dynamics (e.g. traffic changes, link failures), as well as guarantee hard service-level requirements (e.g. ultra-reliable low latency)~\cite{lema2017business}. For example, recent studies in Wide Area Networks show that traffic can change dramatically at the scale of a few minutes ($\approx$1-5 mins)~\cite{wang2021examination,kandula2014calendaring}. This motivates the need for fast network optimizers that can operate at sub-minute timescales.

In this paper, we revisit Traffic Engineering (TE), a fundamental networking task whose goal is to optimize the network performance and resource utilization. TE is a broad topic that comprises a wide spectrum of problems with different optimization goals, and using various traffic control mechanisms (e.g., routing, queue scheduling)~\cite{mendiola2016survey,xu2011link, hartert2015declarative,fortz2000internet}. 
We focus on the classical problem of minimizing the maximum link load in the network (\textit{MinMaxLoad}) by optimizing the configuration of OSPF. This is an NP-hard optimization problem~\cite{xu2011link}.

Given the relevance of this problem, a large body of solutions can be found in the literature~\cite{gay2017repetita,hartert2015declarative,fortz2000internet}. Existing TE solutions can be classified in two main blocks. On the one hand, some solutions rely on general-purpose mathematical programming methods (e.g., using Local Search, Constraint Programming~\cite{gay2017repetita, hartert2015declarative}). These solutions can typically produce near-optimal results with performance and time guarantees. However, they are computationally intensive, which makes them unsuitable for online optimization in modern real-world networks. On the other hand, networking experts have developed over the years efficient handcrafted heuristics for TE~\cite{rexford2006route}. These methods can typically operate faster and scale effectively to large real-world networks. As a counterpart, heuristics typically lead to lower performance compared to the previous TE solvers based on general-purpose mathematical \mbox{programming methods~\cite{gay2017repetita, hartert2015declarative}.}

Our main goal is to design a TE solution that can operate as fast as traditional heuristics, while achieving comparable performance to computationally-intensive TE solvers. In this paper, we present \emph{Routing by Backprop} (\NAME), a novel method for online TE leveraging Graph Neural Networks (GNN)~\cite{battaglia2018relational} and differentiable programming at its core. The code used in our experiments is publicly available, including a short demo\footnote{\url{https://github.com/krzysztofrusek/net2vec}}. Given a network topology and an estimated traffic matrix, \NAME~outputs a set of link weights that optimize the configuration of the OSPF protocol (widely deployed nowadays in networks~\cite{cisco-weights,rexford2006route}). To produce these weights, \NAME~internally implements a GNN-based model that approximates the target TE problem as an end-to-end differentiable function. The resulting GNN-based model introduces a continuous relaxation of the target optimization function. Thus, it is suitable for fast TE via differentiable programming, i.e., by applying a common backpropagation method and optimizing through gradient descent.

In our evaluation, we first apply \NAME~to the \textit{MinMaxLoad} TE problem in four different real-world networks unseen during the training phase. Our results show that ---~by optimizing the OSPF link weights~---  the proposed solution reduces the maximum link load by $\approx$25\% on average, compared to the default OSPF configurations widely deployed by network operators nowadays~\cite{cisco-weights}. At the same time, \NAME~operates at sub-second timescales (median $<$322 ms in networks up to 37 nodes), thus showing promising results for fast TE.

Likewise, we explore the potential of \NAME~to be utilized as an initializer of computationally-intensive TE solvers. In our experiments, we take as a reference IGP-WO, a popular TE solver based on Local Search~\cite{gay2017repetita}. We limit its execution time to only 1 second. As a reference, in~\cite{gay2017repetita} the authors report 5-minute executions to optimize networks of similar size with this TE solver. Our experimental results show that, by initializing IGP-WO with the output produced by \NAME,~we can quickly reduce $\approx$30\% on average the maximum link load with respect to default OSPF settings. This further reveals the potential of the proposed solution to accelerate advanced TE solvers, and thus meet the hard time constraints of online TE.

\section{Proposed Solution: RBB} \label{sec:method}

This section presents the mathematical framework behind Routing by Backprop (RBB). For the mathematical definitions, we treat the network topology as a graph. Then, we use GNNs and soft routing to produce optimized link weights that minimize the maximum link utilization. Lastly, these weights are used by OSPF to update the routing tables on devices, by computing the weighted shortest path (i.e., Dijkstra's algorithm).

Dijkstra’s algorithm is an exact deterministic procedure that maps a weighted graph (e.g., weights on the edges) to a set of shortest paths (i.e., one path for each pair of nodes). This can be seen as a function that maps a vector with weights and a network topology to a set of shortest paths. In our work, we rely on the universal approximation property of neural networks to approximate this function. Specifically, we use GNNs as the universal approximation function, enabling optimization on any continuous function of the shortest path using the gradient with respect to the weights input vector.

\subsection{Notation}

We represent the network as a weighted directed graph $\mathcal G=(\mathcal V,\mathcal E)$, 
where $\mathcal V=\{\bm r_i\}_{i =1:n_v}$ is a set of routers (nodes) and $\mathcal{E}=\{(\bm e_k,r_k,s_k)\}_{k=1:n_e}$ is a set of links (edges) connecting receiver and sender nodes indexed by $r_k,s_k$.
Both, routers and links, are described by their properties (features) respectively denoted as $\bm r_i$ (node features in ML terminology) and $\bm e_{i}$ (edge features).

Routing in the network can be considered as a mapping from a pair of routers to the path connecting them.
Let $u$ and $v$ be the indices of two routers, then the path  can be represented as a binary vector $\bm p_{u,v}\in \{0,1\}^{n_e}$  with 1 on the $k$-th position if the path contains $k$-th link. In this paper we use shortest path routing so these are parameterized by positive link weights (costs) $\bm w = [w_k]_{k=1:n_e}$ such that the path minimizes $\bm p_{u,v}\bm w^T$~\cite{xu2011link}. In this way, the weights indirectly control link load, since each path $\bm p_{u,v}$, has an associated traffic demand volume $d_{u,v}$ routed through it.

Link utilization vector $\bm \rho$ can be expressed with a convenient matrix operation.  Let $\bm u\in \mathbb N^{n_p}$ and $\bm v\in \mathbb N^{n_p}$ be the vectors representing the indices of all possible router pairs ($n_p=n_v(n_v-1)$). Similarly $\bm P \in \{0,1\}^{n_p\times n_e}$ is a matrix that stacks paths for those pairs $\bm P=[\bm p_{u_i,v_i}^T]^T_{i=1:n_p}$.
$\bm P$ allows us to compute the utilization of every link as  $\bm \rho =\bm d \cdot \bm P\odot\frac{1}{\bm c}$, where $\bm d=[d_{u_i,v_i}]_{i=1:n_p}$ and $c_i$ is the capacity of the $i$-th link.
Under common assumption of negligible losess~\cite{fortz2000internet}, the \textit{MinMaxLoad} problem can be described as:
\begin{equation}\label{eq:opt}
	\bm w_{\text{opt}} = \arg\min_{\bm w}\max(\bm\rho(\bm w)).
\end{equation}
RBB proposes a method for relaxing~\eqref{eq:opt} so that a gradient optimization method can be applied.

\subsection{Soft Routing}
\label{subsec:soft_routing}

Differentiable programming was previously discussed in~\cite{liu2020automated} for TE. In that paper, the authors propose to use a deep learning model that makes directly end-to-end TE decisions based on a set of candidate shortest paths. Instead, we consider a soft routing approach based on link weights that makes the gradient more meaningful for gradient-based optimization methods. For this purpose, we use a GNN model that introduces a continuous relaxation on the target step-wise optimization function (where the true gradient is zero in most regions). Moreover, this link-based routing representation is directly compatible with OSPF, which is nowadays the dominant routing protocol in networks~\cite{cisco-weights,rexford2006route}.

We leverage the property that if we set $\bm e_k=[w_k]$ and $\bm r_i=[I(u=i),I(v=i)]$  we end up with the path being a function of graph data ($\bm e_k\in\mathbb R^+, \bm r_i \in\{0,1\}^2$).
Such a function can be approximated by a GNN trained on the routing obtained by the Dijkstra's algorithm (i.e., shortest path).
A GNN is differentiable, however the final discrete path $\bm p_{u,v}$ is not, so we approximate the path by the probabilistic output from the GNN, yet being differentiable:
\begin{equation}\label{eq:gnnapprox}
	\hat{\bm p}_{u,v} = \mathsf{P}(\bm p_{u,v}=1) = f(\mathcal G(\bm w,u,v)).
\end{equation}
Note that~\eqref{eq:gnnapprox} is defined only for a single path.
However, leveraging recent developments in Machine Learning (ML) libraries, we can easily vectorize~\eqref{eq:gnnapprox} into the soft routing matrix $\hat{\bm P}(\bm w) = \textrm{vmap} f(\mathcal G(\bm w,\bm u,\bm v))$ and compute all paths in a single evaluation of batched GNN ---~e.g., using the $\textrm{vmap}$ transform from JAX~\cite{jaxpaper}. From this approximation we get the link utilization as:
\begin{equation}\label{eq:rho}
	\hat{\bm \rho}(\bm w)=\bm d \cdot \hat{\bm P}(\bm w)\odot\frac{1}{\bm c}.
\end{equation}

At this point we could optimize~\eqref{eq:opt}, as the $\max$ function is differentiable. Nevertheless, we observe that using $\textrm{softmax}$ can be beneficial to achieve more relaxation on the target optimization goal. 
In particular, we use $\textrm{softmax}$ with temperature $\tau$ that controls its steepness (relaxation) to approximate the max function:
\begin{equation}\label{eq:softmaximum}
	\widetilde{\max}(\bm x)\equiv \frac{\sum_i x_ie^\frac{x_i}{\tau}}{\sum_j e^\frac{x_j}{\tau}}.
\end{equation}

Combining~\eqref{eq:rho} with~\eqref{eq:softmaximum} in the gradient descent algorithm, we propose the following weight update rule that minimizes the maximum utilization:
 \begin{equation}\label{eq:gradientdescent}
 	\bm w_{t+1}=\bm w_{t} - \alpha_t \nabla \widetilde{\max}\left (\hat{\bm \rho}(\bm w_t) \right),
 \end{equation}
where $\alpha_t>0$ is the learning rate.
Here we rely on automatic differentiation to compute $\nabla \widetilde{\max}\left (\hat{\bm \rho}(\bm w_t) \right)$~\cite{autodiff}. Note that we are computing gradients with respect to the model inputs instead of the internal GNN weights, so we cannot reuse the gradients used for training the GNN model.

\subsection{Graph Neural Network}
\label{subsec:gnn}

In the proposed method, the only trainable part is the relaxed shortest path calculation from link weights in~\eqref{eq:gnnapprox}, which is implemented by a GNN. Further calculations are based on prior knowledge from existing analytical network models~\cite{rexford2006route}. Below we elaborate on the design of our GNN model to approximate~\eqref{eq:gnnapprox}. We refer the reader to~\cite{battaglia2018relational} for generic background on GNNs.

We adopt a highly flexible architecture named encode-process-decode~\cite{battaglia2018relational}.
Such an architecture consists of an encoder model that independently encodes node and edge attributes ($\bm e_i$ and $\bm r_i$).
A core model, which performs $T$ rounds of message-passing steps and 
a decoder that independently decodes the resulting node and edge attributes.

The encoder and decoder in our proposal are implemented with 2-layer dense neural networks followed by a normalization layer. The processor consists of $T$ graph network blocks~\cite{battaglia2018relational}. Each block updates the edge representation by applying a dense neural network $f^e$ (similar to the encoder) to the concatenation of the edge, sender, and receiver node attributes: $\bm e_k^{t+1}=f^e([\bm e_k^t, \bm r_{r_k}^t,\bm r_{s_k}^t])$. The attributes of all edges adjacent to a given node are aggregated by a function, which in our case is a summation: $\bar{\bm e}_i^t=\sum_{k:r_k=i} \bm e_k^t$. Given the information from the neighboring edges, the node representation is updated by another neural network $f^v$: $\bm r_{i}^{t+1}=f^v([\bar{\bm e}_i^t,\bm r_{i}^{t}])$.

The final output of the GNN is the result of the decoder network being independently applied to all node and edge attributes. In our case, we only generate edge-level outputs, which represent the probability that a given link belongs to a shortest path. The resulting GNN was trained by backpropagation using cross entropy as a loss function between predictions and the true values from the Dijkstra's algorithm. We treat it as a classification problem. Particularly, we use the loss computed at each message-passing step and backpropagate the sum. In this way, each graph network block learns how to iteratively improve upon the previous one, which introduces a bias that may help with faster convergence~\cite{battaglia2018relational}.

Lastly, we combine this GNN model with the soft routing approach of Section~\ref{subsec:soft_routing}. Particularly, the GNN model acts as a differentiable surrogate method for $\hat{\bm P}(\bm w)$ in Equation~\eqref{eq:rho}. This makes it possible to build an end-to-end differentiable function that models the target TE problem (i.e., MinMaxLoad).

\section{Evaluation}

In this section, we evaluate RBB under a wide variety of network topologies and traffic matrices. We also test RBB's potential for accelerating high-performance TE solvers based on general-purpose mathematical programming.

In the experiments we consider that all links have a capacity of 1 Mbps and traffic matrices are generated using a uniform distribution. Specifically, for each node pair, the traffic generated is uniformly sampled from the interval (0, \textit{n-1}), with \textit{n} being the number of nodes in the graph. The resulting traffic is then injected into the network following the shortest path routing (i.e., using the link-level weights). We scale the traffic matrix values by a topology-specific number in such a way that the link load distribution has a soft tail above 1 Mbps (i.e., there are a few overloaded links).

\subsection{Training of the GNN model}
To train the GNN model, we first generate a dataset with random topology graphs using the Barab\'{a}si-Albert (BA) model~\cite{Barabasi1999}. The number of nodes was randomly sampled between the range $[10,20]$. In all topologies we sampled the link weights from a scaled beta distribution in the range $(0,2)$ with mode~1. For each training sample, we selected two random points (src,~dst) and computed the weighted shortest path between them using Dijkstra's algorithm.

All components of the encode-process-decode GNN architecture were implemented with a 2-layer MLP with 128 units, each followed by a normalization layer. The output of the decode function uses a sigmoid activation to produce link probabilities in the range $(0,1)$. The model was trained using Adam optimizer with learning rate of 0.001 and with batches of 32 graphs. For the validation, we considered 2,000 samples of 4 real-world network topologies extracted from SNDLib~\cite{orlowski2010sndlib} and described in Table~\ref{tab:summary}.

 We observed loss convergence after 8,000 training steps. In our setup, the whole training process took $\approx$51 min. At this point, the model achieved a link prediction accuracy of 0.9893 $\pm$ 0.0005 on the validation samples. Once the model was trained, we used it as a differentiable surrogate for shortest path routing to implement the optimization method described in Section~\ref{sec:method}.

\subsection{Comparison with Default OSPF}
\label{subsec:imp_ospf}

\begin{table}[t]
    \centering
        \caption{Percentage of cases where RBB outperforms Default OSPF}
    \label{tab:summary}
\begin{tabular}{l|ll|c}
\textbf{Network} & \textbf{$n_v$ (nodes)} & \textbf{$n_e$ (links)} & \textbf{$\overline{\max\bm\rho_0>\max\bm\rho_3}$}\\
\toprule
cost266 & 37 & 57 & 100\%\\
geant & 22 & 36 & 99.4\%\\
janos-us & 26 & 84 & 100\%\\
nobel-germany & 17 & 26 & 99.9\%\\
\end{tabular}
\end{table}

\begin{figure}[t]
    \centering
    \includegraphics{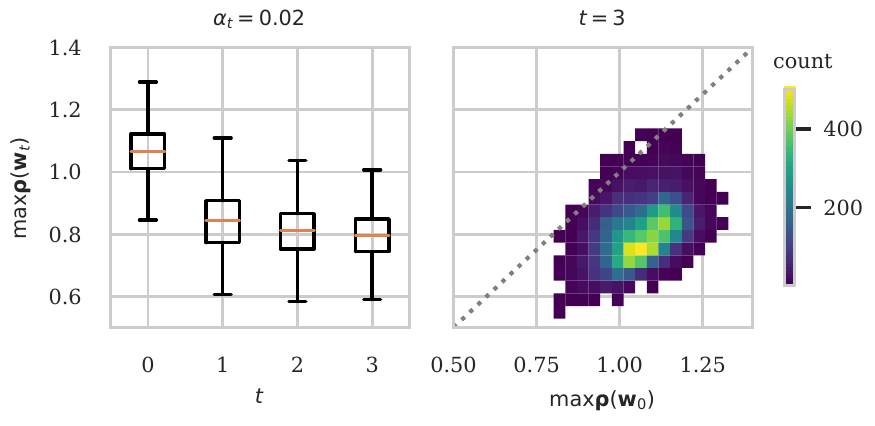}

 \caption{Comparison with \textit{Default OSPF} in 12,000 samples from the 4 real-world topologies of Table~\ref{tab:summary}. Left:~Evolution of the maximum link utilization (y-axis) per optimization step with RBB (x-axis). $t$$=$$0$ represents the results of the initial \textit{Default OSPF} configuration. Right: Joint histogram of the maximum link utilization. X-axis represents the results of \textit{Default OSPF}, and the y-axis shows the results after three optimization steps with RBB. }
 \label{fig:imprv_ospf}
\end{figure}

In this section, we explore the capabilities of \NAME~to optimize the routing configuration in OSPF. Specifically, \NAME~produces new link weights according to Equation~\eqref{eq:gradientdescent} so that the maximum link utilization is minimized. For each experiment, we considered a topology from Table~\ref{tab:summary} where the traffic matrix  was sampled from a uniform distribution. We initialized the link weights inversely proportional to the link capacity, which is a default configuration commonly adopted by operators~\cite{cisco-weights}. Hereafter, we refer to this baseline as \textit{Default OSPF}. In the optimizer, we set $\tau = 0.1$ for the softmax function in Equation~\eqref{eq:softmaximum}, and $\alpha=0.02$ in Equation~\eqref{eq:gradientdescent}.

We evaluated \NAME~on 12,000 samples from the 4 real-world topologies in Table~\ref{tab:summary} (i.e., 3,000 samples from each topology with different traffic matrices). Figure~\ref{fig:imprv_ospf} (left) shows the results for the first three optimization steps ($t$$=$$0$ corresponds to the initial \textit{Default OSPF} configuration). The results indicate that it takes only one optimization step ($t$$=$$1$) to reduce the maximum link utilization below 1 in over 93\% of the cases. Likewise, after three optimization steps RBB avoids overloaded links (i.e., utilization$<$1) in 99.8\% of the cases. On average, the maximum link utilization is reduced by \textbf{$\approx$25\%} with respect to \textit{Default OSPF} after three optimization steps ($t$$=$$3$).
 
In Figure~\ref{fig:imprv_ospf} (right) we observe the joint histogram of initial and final maximum link utilizations after three optimization steps. Most of the experiments are far below the diagonal line, indicating that the final configuration produced by \NAME~considerably outperforms \textit{Default OSPF}. Table~\ref{tab:summary} (right column) shows the portion of cases where \NAME~outperforms \textit{Default OSPF} in each network. It can happen ---~although rarely (\mbox{$<0.6\%$})~--- that the final configuration is worse than the initial one. Nevertheless, in a practical application the final configurations could be quickly verified before deployment so that in those rare cases \textit{Default OSPF} could be used instead.

\subsection{Execution Time}

\begin{table}[b]
    \centering
        \caption{Execution times of an optimization step \\(median values over 3,000 runs)}
    \label{tab:execution-cost}
\begin{tabular}{l|cc}
\textbf{Network} & \textbf{Execution time} [ms]\\
\toprule
cost266 & 322\\
geant & 76\\
janos-us & 122\\
nobel-germany & 36 \\
\end{tabular}
\end{table}

In the proposed method, the GNN model is evaluated for each source-destination pair to account for the whole routing, so the numerical complexity is quadratic $O(n_v^2)$. To obtain the highest possible performance we leveraged the JAX library that offers automatic differentiation, vectorization, and a JIT compiler based on XLA~\cite{jaxpaper}.

First, the GNN is vectorized to compute all paths at once. Then, the model is transformed to compute the gradient of the softmax of the link utilization with respect to the link weights.
Finally, the whole operation is compiled for a GPU (Tesla V100-SXM2-32GB in our experiments). Table~\ref{tab:execution-cost} shows the resulting execution times for one optimization step (median of 3,000 repetitions, confidence interval is less than the least significant digit). For larger topologies, it would be possible to distribute the computation of different source-destination pairs across multiple GPUs. 

\subsection{Bootstrapping high-performance TE solvers}

\begin{figure}[t]
    \centering
    \includegraphics[width=1.0\linewidth]{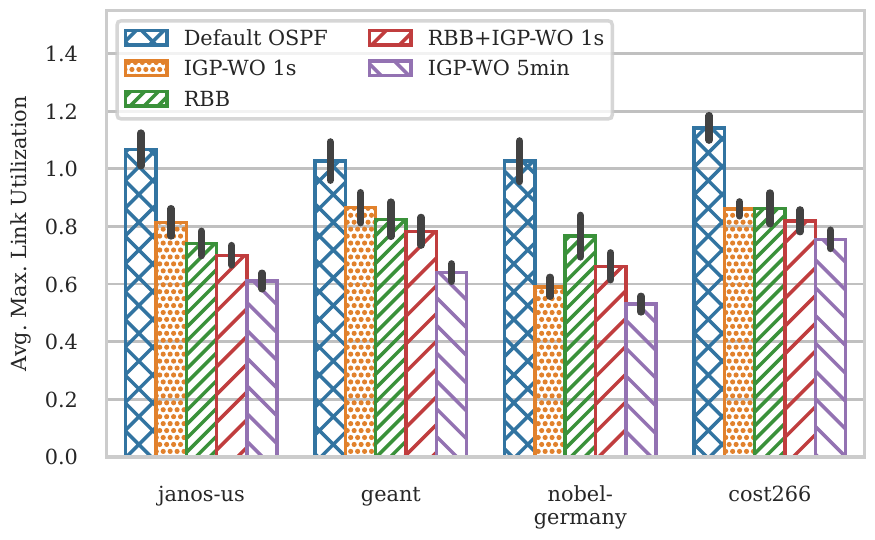}
    \caption{Comparison between Default OSPF, RBB and IGP-WO in 4 real-world topologies (each with 3,000 samples with different traffic matrices).}
    \label{fig:barplot}
\end{figure}

In this section we test the potential of \NAME~to act as an initializer to other computationally-intensive TE methods. To this end, we combined \NAME~with a state-of-the-art link weight optimization algorithm called \textit{IGP-WO}~\cite{gay2017repetita}, which is based on Local Search. A main advantage of IGP-WO is that it is an anytime optimization method, meaning that the optimization process can be stopped at any moment and the result obtained is valid~\cite{fortz2000internet}.

Similarly to Section~\ref{subsec:imp_ospf}, we executed three optimization steps with the \NAME~method. The resulting link weights were used to initialize the IGP-WO method (hereafter, referred to as \mbox{\textit{RBB+IGP-WO}}). To test the potential of this solution for online TE, we limited the execution of the IGP-WO method to only 1 second, and we compared the results against the standard IGP-WO implementation, which was initialized with \textit{Default OSPF}. As a reference, we also evaluated the results with 5-minute executions of standard IGP-WO, which represents the times reported in~\cite{gay2017repetita} to achieve near-optimal results in networks of similar size to those of our experiments.

Figure~\ref{fig:barplot} shows the maximum link utilization achieved by the different optimization methods for samples of different topologies. We observe that by initializing IGP-WO with \NAME~and executing it only for 1 second it reduces by $\approx$30\% on average the maximum link load with respect to \textit{Default OSPF}. In general, we can see that the \textit{RBB+IGP-WO} method significantly outperforms the standard IGP-WO method with 1-second executions. However, in samples of the \textit{nobel-germany} topology the performance is slightly degraded. 
This is the smallest topology we tested. As a result, it has a narrower space of near-optimal routing configurations compared to the other topologies. This may partially explain why the algorithm performs worse here. We leave outside the scope of this paper a deeper analysis on how the topology characteristics may affect the performance of \NAME.  

The experimental results suggest that \NAME~may serve as a good initialization method for computationally-intensive TE solvers. In this case, the \NAME~method would only introduce a limited execution overhead at the scale of sub-seconds (see execution times in Table~\ref{tab:execution-cost}).

\section{Discussion}

The main difficulty in TE problems comes from the discrete nature of the cost functions. A potential limitation of RBB is that it optimizes the continuous surrogate cost, which may not approximate accurately the true discrete function. For this purpose, we control the level of approximation by tuning some parameters of our soft approximation to discrete functions (e.g., the temperature). 

From a practical standpoint, RBB optimizes the link weights at the cost of executing an intensive training process. This means that it needs to spend some time to be trained before being deployed for network optimization. Once trained, the optimization time is faster than other iterative link weight optimization solutions (e.g., \cite{fortz2000internet, 9651930}). The generalization capabilities of GNNs enable RBB to achieve a high optimization performance for a wide range of network scenarios not seen during the training process.

\section{Conclusion}

In this paper we presented \textit{Routing by Backprop} (RBB), a novel method that leverages differential programming for fast TE optimization. \NAME~implements internally a GNN model that allows for the continuous relaxation of the optimization function (\textit{MinMaxLoad} in our case). This eventually permits fast optimization using a gradient descent method. Our experimental results show that \NAME~considerably outperforms default OSPF configurations widely deployed nowadays by network operators ($\approx$25\% of improvement in our case). Moreover, we show that the proposed ML-based method generalizes successfully over different network topologies and traffic matrices not seen during training.

As future work, we plan to extend the current GNN model to support Equal Cost Multi-Path (ECMP), which is widely supported nowadays in networks running OSPF. Also, we plan to extend RBB to support the optimization of more complex performance metrics (e.g., end-to-end delay, jitter), as well as to perform multi-objective optimization.

\section*{Acknowledgment}

This work was supported by the Polish Ministry of Science and Higher Education with the subvention funds of the Faculty of Computer Science, Electronics and Telecommunications of AGH University and by the PL-Grid Infrastructure. Also, this publication is part of the
 Spanish I+D+i project TRAINER-A (ref.~PID2020-118011GB-C21), funded by MCIN/ AEI/10.13039/501100011033. This work is also partially funded by the Catalan Institution for Research and Advanced Studies (ICREA) and the Secretariat for Universities and Research of the Ministry of Business and Knowledge of the Government of Catalonia and the European Social Fund.

\bibliographystyle{IEEEtran}
\bibliography{ref.bib}

\end{document}